\documentstyle[12pt,amsmath,amsfonts, graphicx]{article}

\numberwithin{equation}{section}

\begin{document}

\hoffset = -1truecm
\voffset = -2truecm

\begin{titlepage}
{\flushright
\today
\\}
\vskip 2truecm {\center\LARGE \bf BPS Domain Walls and Vacuum Structure of $N=1$
Supergravity Coupled To A Chiral Multiplet
\\} \vskip 2truecm

{\center \bf
 Bobby E. Gunara$^{\flat}$, Freddy P. Zen$^{\flat}$, and Arianto$^{\flat,\sharp}$
 \footnote{email: bobby@fi.itb.ac.id, fpzen@fi.itb.ac.id, feranie@upi.edu}
\\}

\vskip 1truecm

\begin{center}
$^{\flat}$\textit{Theoretical  Physics Laboratory, \\
Theoretical High Energy Physics and Instrumentation Research
Group,\\
Faculty of Mathematics and Natural Sciences, \\
Bandung Institute of Technology \\
Jl. Ganesha 10 Bandung 40132, Indonesia.}

 \vskip 0.5truecm

$^{\sharp}$\textit{Department of Physics, \\
Faculty of Mathematics and Natural Sciences, \\
Udayana University \\
Jl. Kampus Bukit Jimbaran-Kuta Denpasar 80361, Indonesia.}

\end{center}

\vskip 2truecm

{\center \large \bf
ABSTRACT
\\}
\vskip 1truecm

\noindent We study BPS domain walls of $N=1$ supergravity coupled to a chiral
 multiplet and  their Lorentz invariant vacuums  which
can be viewed as critical points of BPS equations and the scalar
potential. Supersymmetry further implies that gradient flows of
BPS equations controlled by a holomorphic superpotential and the
K\"ahler geometry are unstable near local maximum of the scalar
potential, whereas they are stable around local minimum and
saddles of the scalar potential. However, the analysis using
renormalization group flows shows that such gradient flows do not
always exist particularly in infrared region.

\end{titlepage}




\section{Introduction}
Topological defects such as domain wall solutions of supergravity
have acquired a large interest due to their duality with
renormalization group (RG) flows described by a beta function
 of field theory in the context of AdS/CFT correspondence \cite{AdSCFT}. In particular
 there has been  a lot of study considering these solutions which preserve some fraction
 of supersymmetry  in five dimensional supergravity theory, see, for example, in Ref.~\cite{AC}.
 However, apart  from this application our basic interest in domain walls is to investigate
 general properties  of supersymmetric flows, namely gradient flows and RG flows, and
 supersymmetric  Lorentz invariant vacuums of four dimensional $N=1$ supergravity coupled
 to a chiral multiplet. The walls preserve half of supersymmetry which are called BPS walls
 and were first discovered in Ref. \cite{BPSDW}\footnote{BPS domain walls of four dimensional
 $N=1$ supergravity were extensively reviewed in Ref. \cite{RevBPSDW}. We thank M. Cvetic for pointing out
 these references.}.\\
\indent The purpose of this paper is to present our study on BPS
domain walls describing partial Lorentz invariant ground states of
$N=1$ supergravity coupled to a chiral multiplet in four
dimensions and properties of their $N=1$ critical points
(equilibrium points) representing supersymmetric Lorentz invariant
ground states, namely, Minkowskian and AdS ground states. The
walls preserve half of supersymmetry and are described by BPS
equations which can be seen as a set of two dimensional autonomous
dynamical system\footnote{The analysis using dynamical system has
also been discussed for dilaton domain walls in Ref. \cite{DDW}.}.
Moreover, stability of gradient flows described by BPS equations
is controlled by a holomorphic superpotential and
 the K\"ahler geometry.\\
\indent Our first order analysis of BPS equations around critical points shows
 that in this model the gradient flows are unstable near local maximum of
the scalar potential, whereas they are stable near saddles, local minimum, and degenerate
critical points of the scalar potential. However, it turns out that such technique does not
always work in general. First, the procedure fails when the flows become degenerate which
occur near \textit{intrinsic} degenerate critical points of the scalar potential\footnote{We distinguish
 the notion of \textit{intrinsic} degenerate critical points with degenerate critical points
 of the scalar potential, see section \ref{CPWV} for detail.}.
Second, to get a complete picture one has to perform first order
analysis on beta function. This analysis leads us to conclude that
if parameters in the superpotential vary, in a region there is a
possibility of having a zero eigenvalue in the analysis, then no
critical points of the scalar potential exist in this region which
follows that any gradient flow does not exist .
For the case at hand, such situation occurs in infrared (low energy scale) region.\\
\indent The organization of this paper is as follows. In section
\ref{BPSDW} we provide a quick review of four dimensional $N=1$
supergravity coupled to a chiral multiplet and also their
supersymmetric solitonic solution BPS domain walls. In section
\ref{CPWV} we consider properties of critical points describing
supersymmetric ground states of BPS equations and scalar potential
which are useful for our analysis in the next sections. Then we
discuss the nature of BPS equation and beta function describing
gradient and RG flows, respectively in
section \ref{GRG}. Application of our analysis in several models
is set in section \ref{Models}. Finally we conclude our results in
section {\ref{Conclu}}.

\section{BPS Domain Walls of $4d\: N=1$ Supergravity Coupled to A Chiral Multiplet}
\label{BPSDW}

This section is devoted to give a short review of four dimensional
$N=1$ supergravity coupled to a chiral multiplet and further
discuss its solitonic solutions called domain walls which do not
preserve fully Lorentz invariance. In particular, we only consider
domain walls which inherit half of supersymmetry of the parental
theory, \textit{i.e.}, the four dimensional $N=1$ supergravity.
Such solution is called BPS domain wall \footnote{In this section
some results have been discussed in Ref.~\cite{CDGKL} in the
context of stringy  landscape. We thank K. Yamamoto for
noticing us this paper.}.\\
\indent Let us first discuss the $N=1$ supergravity coupled to a
chiral multiplet. This theory consists of a gravitational
multiplet and a chiral multiplet. The gravitational multiplet
contains a vierbein $e^a_\mu$ and a vector fermion $\psi_\mu$
where $a=0,...,3$ and $\mu=0,...,3$ are the flat and the curved
indices, respectively. A chiral multiplet is composed of a complex
scalar $z$
and a spin-$\small{\frac{1}{2}}$ fermion $\chi$.\\
\indent The complete $N=1$ supergravity Lagrangian together with its supersymmetry transformation
can be found, for example, in Ref.~\cite{susy}. Here we collect the terms which are useful for our analysis.
The bosonic part of the $N=1$ supergravity Lagrangian can be written as
\begin{equation}
{\mathcal{L}}^{N=1} = -\frac{1}{2M^2_P}R +
g_{z\bar{z}}\partial_{\mu} z
\partial^{\mu}\bar{z} - V(z,\bar{z})\quad, \label{L1}
\end{equation}
where the complex scalars $(z,\bar{z})$ span a Hodge-K\"ahler
manifold with metric $g_{z\bar{z}}= \partial_{z}
\partial_{\bar{z}}K(z,\bar{z})$ and $K(z,\bar{z})$ is a real
function called K\"ahler potential. The $N=1$ scalar potential
$V(z,\bar{z})$ has the form
\begin{equation}
 V(z,\bar{z}) = e^{K/M^2_P}\left(g^{z\bar{z}}\nabla_z W\,\bar{\nabla}_{\bar{z}} \bar{W}
 - \frac{3}{M^2_P} W \bar{W} \right)\quad,
\label{V}
\end{equation}
where $W$ is a holomorphic superpotential and $\nabla_z
W\equiv (dW/dz)+ (K_z/M^2_P) W$. The supersymmetry
transformations up to three-fermion terms leaving invariant (\ref{L1})
are
\begin{eqnarray}
\delta\psi_{1\mu} &=& \frac{1}{M_P}D_\mu\epsilon_1
+ \frac{\mathrm{i}}{2M_P}e^{K/2M^2_P}\,W\gamma_\mu \epsilon^1 \quad, \nonumber\\
\delta\chi^z &=& {\mathrm{i}}\partial_\mu z \, \gamma^\mu\epsilon^1 + N^z\epsilon_1 \quad, \label{susytr}\\
\delta e^a_\mu &=&  -{\mathrm{i}} M_P ( \bar{\psi}_{1\mu} \,
\gamma^a \epsilon^1
+ \bar{\psi}^1_\mu \, \gamma^a \epsilon_1 )\quad,\nonumber\\
\delta z &=& \bar{\chi}^z \epsilon_1 \quad,\nonumber
\end{eqnarray}
where $N^z \equiv
e^{K/2M^2_P}\,g^{z\bar{z}}\bar{\nabla}_{\bar{z}}\bar{W}$
and $g^{z\bar{z}} = (g_{z\bar{z}})^{-1}$. The coupling constant $M_P$ is the Planck mass that
one regains $N=1$ global supersymmetric theory by setting $M_P \to +\infty$.\\
\indent Now we will discuss the ground states which break
partially Lorentz invariance, \textit{i.e.}, domain walls. First
of all, one has to take the ansatz metric as
\begin{equation}
ds^2 =
a^2(u)\,\eta_{\overline{\mu}\,\overline{\nu}}\,dx^{\overline{\mu}}dx^{\overline{\nu}}-du^2
\quad,\label{DWans}
\end{equation}
where $\overline{\mu},\,\overline{\nu}=0,1,2$ and $a(u)$ is the
warped factor. Writing the supersymmetry transformation
(\ref{susytr}) and setting $\psi_{1\mu}=\chi^z=0$ on the
background (\ref{DWans}), this leads
\begin{eqnarray}
\delta\psi_{1u} &=& \frac{1}{M_P}D_u \,\epsilon_1 + \frac{\mathrm{i}}{2M_P}e^{K/2M^2_P}\,
W\gamma_u \epsilon^1 \quad, \nonumber\\
\delta\psi_{1\overline{\mu}} &=& \frac{1}{M_P}\partial_{\overline{\mu}}\,\epsilon_1
+ \frac{1}{2M_P}\gamma_{\overline{\mu}}\left(-\frac{a'}{a}\gamma_3 \epsilon_1
+ {\mathrm{i}}e^{K/2M_P^2}\, W \epsilon^1 \right) +... \quad, \label{susytr1}\\
\delta\chi^z &=& {\mathrm{i}}\partial_\mu z \,
\gamma^\mu\epsilon^1 + N^z\epsilon_1 \quad, \nonumber
\end{eqnarray}
where $a' \equiv da/du$ and the dots mean that these term can be
neglected in our analysis. Furthermore, in order to have residual
supersymmetry on the ground states, the equations in Eq.
(\ref{susytr1}) have to be vanished. For the sake of simplicity,
we assign that $\epsilon_1$ and $z$ are only $u$ dependent. Thus,
the first equation in Eq. (\ref{susytr1}) shows that $\epsilon_1$
depends only on $u$, while the second equation gives a projection
equation
\begin{equation}
\frac{a'}{a}\gamma_3 \epsilon_1  = {\mathrm{i}}e^{K/2M^2_P}\, W(z)
\epsilon^1 \quad,\label{projector}
\end{equation}
which leads to
\begin{equation}
\frac{a'}{a}  = \pm \,e^{K/2M^2_P}\, \vert W(z) \vert \quad.
\label{warp}
\end{equation}
The third equation in Eq.~(\ref{susytr1}) becomes
\begin{eqnarray}
z' &=& \mp 2 g^{z\bar{z}}
\bar{\partial}_{\bar{z}}{\mathcal{W}}(z,\bar{z}) \quad,\nonumber\\
\bar{z}' &=& \mp 2 g^{z\bar{z}}
\partial_z{\mathcal{W}}(z,\bar{z}) \quad,
\label{gfe}
\end{eqnarray}
where we have introduced a real function ${\mathcal{W}}(z,\bar{z})
\equiv e^{K/2M^2_P}\, \lvert W(z) \rvert$. The gradient flow equations (\ref{gfe}) are
called BPS equations. Note that using Eq.~(\ref{gfe}) one shows that the function (\ref{warp})
is monotonically decreasing and related to the monotonicity of the c-function in the holographic
correspondence \cite{DWc}. Furthermore, from conformal field theory (CFT) point of view the relevant
 supersymmetric flows for all scalar fields can also be described by a beta function defined as
\begin{equation}
 \beta \equiv a \frac{dz}{da} = - 2g^{z \bar{z}} \frac{\bar{\partial}_{\bar{z}} {\mathcal{W}}}{\mathcal{W}} \quad,
\label{beta}
\end{equation}
 after using Eqs.~(\ref{warp}) and (\ref{gfe}), together with its complex conjugate. In this picture
 the scalars can be viewed  as coupling constants and the warp factor $a$ is playing the role
 of an energy scale \cite{DWc, CDKV}. \\
\indent Now  the potential (\ref{V}) can be cast into the form
\begin{equation}
 V(z,\bar{z}) = 4\, g^{z\bar{z}}\, \partial_z {\mathcal{W}}\,
 \bar{\partial}_{\bar{z}} {\mathcal{W}}
 - \frac{3}{M^2_P}\, {\mathcal{W}}^2 \quad,
\label{V1}
\end{equation}
and furthermore its first derivative with respect to $(z,\bar{z})$
is given by
\begin{eqnarray}
\frac{\partial V}{\partial z} &=& 4\, g^{z\bar{z}}\,
\nabla_z\partial_z {\mathcal{W}}\, \bar{\partial}_{\bar{z}}
{\mathcal{W}}+ 4\, g^{z\bar{z}}\, \partial_z {\mathcal{W}}\,
\partial_z \bar{\partial}_{\bar{z}} {\mathcal{W}}
 - \frac{6}{M^2_P}\, {\mathcal{W}}\,\partial_z{\mathcal{W}}\quad ,\nonumber\\
\frac{\partial V}{\partial \bar{z}} &=& 4\, g^{z\bar{z}}\,
\bar{\nabla}_{\bar{z}}\bar{\partial}_{\bar{z}}{\mathcal{W}}\,
\partial_z{\mathcal{W}}+ 4\, g^{z\bar{z}}\,
\bar{\partial}_{\bar{z}}{\mathcal{W}}\, \bar{\partial}_{\bar{z}}
\partial_z {\mathcal{W}}
 - \frac{6}{M^2_P}\, {\mathcal{W}}\,\bar{\partial}_{\bar{z}}{\mathcal{W}}\quad
 ,\label{dV}
\end{eqnarray}
where $\nabla_z\partial_z {\mathcal{W}}= \partial^2_z
{\mathcal{W}} - \Gamma^z_{zz}\partial_z {\mathcal{W}}$.\\
 \indent We give now our attention to the gradient flow equation (\ref{gfe}) and the first
 derivative of the scalar potential (\ref{dV}). Critical points of
 Eq.~(\ref{gfe}) are related to the following condition
\begin{equation}
\partial_z{\mathcal{W}}(z,\bar{z}) = \bar{\partial}_{\bar{z}}{\mathcal{W}}(z,\bar{z})
=0 \quad,
\end{equation}
which implies that
\begin{equation}
\frac{\partial V}{\partial z} = \frac{\partial V}{\partial
\bar{z}} =0 \quad.
\end{equation}
This means that the critical points of ${\mathcal{W}}(z,\bar{z})$
are somehow related to the critical points of the $N=1$ scalar
potential $V(z,\bar{z})$. Moreover, in the view of Eq.~(\ref{beta}), these
points are in ultraviolet (UV) region if $a \to \infty$ and in infrared (IR) region
if $a \to 0$. Thus, the RG flow interpolates between UV and IR critical points.
 These will be discussed in section \ref{GRG}.

\section{Critical Points of ${\mathcal{W}}(z,\bar{z})$ and $V(z,\bar{z})$}
\label{CPWV}

In this section our discussion will be focused on general properties
of critical points of the real function ${\mathcal{W}}(z,\bar{z})$
and the $N=1$ scalar potential $V(z,\bar{z})$ describing supersymmetric
Lorentz invariant vacuums\footnote{In the
paper we assume ${\mathcal{W}}(z,\bar{z})$ and $V(z,\bar{z})$ to
be $C^\infty$-function.}. General theory of critical points of
surfaces can be found in, for example, Eq.~\cite{calculus} and Appendix \ref{Hess}.\\
\indent First of all we discuss the critical points of
${\mathcal{W}}(z,\bar{z})$. It is straightforward to write down
the eigenvalues and the determinant of the Hessian matrix of
${\mathcal{W}}(z,\bar{z})$ evaluated at the critical point $p_0
\equiv (z_0,\bar{z}_0)$,
\begin{eqnarray}
 \lambda^{\mathcal{W}}_{1,2} &=& \frac{g_{z\bar{z}}(p_0)}{M_P^2}{\mathcal{W}}(p_0)
 \pm 2 \lvert \partial^2_z {\mathcal{W}}(p_0)\rvert \nonumber \:,\\
 {\textrm{det}}H_{\mathcal{W}} &=&  \frac{g^2_{z\bar{z}}(p_0)}{M_P^4}{\mathcal{W}}^2(p_0)
 - 4 \lvert \partial^2_z {\mathcal{W}}(p_0)\rvert^2 \:,
\label{eigenvalW}
\end{eqnarray}
where
\begin{equation}
\partial^2_z {\mathcal{W}}(p_0) = \frac{e^{K (p_0)/M^2_P}\,\bar{W}(\bar{z}_0)}{2{\mathcal{W}}(p_0)}
\left( \frac{d^2W}{dz^2}(z_0) + \frac{K_{zz}(p_0)}{M_P^2}W(z_0)+ \frac{K_z(p_0)}{M_P^2}\frac{dW}{dz}(z_0)\right)\:.
\label{duaW}
\end{equation}
Local minimum point occurs if
\begin{equation}
   \lvert \partial^2_z {\mathcal{W}}(p_0)\rvert < \frac{1}{2M_P^2}g_{z\bar{z}}(p_0){\mathcal{W}}(p_0)\:,
\label{minW}
\end{equation}
while saddle point requires
\begin{equation}
   \lvert \partial^2_z {\mathcal{W}}(p_0)\rvert > \frac{1}{2M_P^2}g_{z\bar{z}}(p_0){\mathcal{W}}(p_0)
   \:.\label{saddW}
\end{equation}
Moreover, critical point becomes degenerate if
\begin{equation}
   \lvert \partial^2_z {\mathcal{W}}(p_0)\rvert = \frac{1}{2M_P^2}g_{z\bar{z}}(p_0){\mathcal{W}}(p_0)
   \:,\label{degW}
\end{equation}
and in general there is no local maximum point for the model.\\
\indent Now we investigate critical points of the scalar potential
$V(z,\bar{z})$ and its relation to the critical points of
${\mathcal{W}}(z,\bar{z})$. The eigenvalues and the determinant of
the Hessian matrix of  $V(z,\bar{z})$ evaluated at the critical
point $p_0$ can be expressed as follows,
\begin{eqnarray}
 \lambda^{V}_{1,2} &=& -2 \left(g^{z\bar{z}}(p_0)\, {\textrm{det}}H_{\mathcal{W}}
 + \frac{g_{z\bar{z}}(p_0)}{M_P^4}{\mathcal{W}}^2(p_0)\right) \nonumber\\
 && \pm \, 2 \frac{{\mathcal{W}}(p_0)}{M_P^2}  \: \left\lbrack \frac{g^2_{z\bar{z}}(p_0)}{M_P^4}{\mathcal{W}}^2(p_0)
 -{\textrm{det}}H_{\mathcal{W}} \right\rbrack ^{1 \over 2}   \:,\label{eigenvalV}\\
&& \nonumber\\
 {\textrm{det}}H_V &=& 4 \,{\textrm{det}}H_{\mathcal{W}}\:\left\lbrack g^{-2}_{z\bar{z}}(p_0){
 \textrm{det}}H_{\mathcal{W}} + \frac{3 {\mathcal{W}}^2(p_0)}{M_P^4} \right\rbrack \nonumber\:.
\end{eqnarray}

We find that local minimum of the scalar potential $V(z,\bar{z})$
has to be fulfilled
\begin{equation}
   \lvert \partial^2_z {\mathcal{W}}(p_0)\rvert > \frac{g_{z\bar{z}}(p_0)}{M_P^2}{\mathcal{W}}(p_0)    \:,
   \label{minV}
\end{equation}
whereas local maximum satisfies
\begin{equation}
   \lvert \partial^2_z {\mathcal{W}}(p_0)\rvert < \frac{g_{z\bar{z}}(p_0)}{2M_P^2}{\mathcal{W}}(p_0)
   \:.\label{maxV}
\end{equation}
Furthermore saddle point occurs if
\begin{equation}
  \frac{g_{z\bar{z}}(p_0)}{2M_P^2}{\mathcal{W}}(p_0)  < \lvert \partial^2_z {\mathcal{W}}(p_0)\rvert
  < \frac{g_{z\bar{z}}(p_0)}{M_P^2}{\mathcal{W}}(p_0)
  \:.\label{saddV}
\end{equation}
There is a possibility of having degenerate critical points of $V(z,\bar{z})$, namely,
\begin{eqnarray}
  \lvert \partial^2_z {\mathcal{W}}(p_0)\rvert &=&
\frac{g_{z\bar{z}}(p_0)}{2M_P^2}{\mathcal{W}}(p_0) \;, \nonumber\\
\lvert \partial^2_z {\mathcal{W}}(p_0)\rvert &=&
\frac{g_{z\bar{z}}(p_0)}{M_P^2}{\mathcal{W}}(p_0)
  \:.\label{degV}
\end{eqnarray}
Our comments are in order. From equations (\ref{minW})-(\ref{saddW}) and (\ref{minV})-(\ref{saddV})
we find that any local minima of ${\mathcal{W}}(z,\bar{z})$ are mapped into local
maxima of the scalar potential $V(z,\bar{z})$. On the other side,  saddles of ${\mathcal{W}}(z,\bar{z})$
 are mapped into  saddles or local minima of the scalar potential $V(z,\bar{z})$.
Moreover the first equation in Eq. (\ref{degV}) comes naturally
from the fact that degenerate critical points of
${\mathcal{W}}(z,\bar{z})$, \textit{i.e.}, Eq. (\ref{degW}),  are
mapped into degenerate critical points of $V(z,\bar{z})$. In this
case, both Hessian matrix $H_{\mathcal{W}}$ and $H_{\mathcal{V}}$
are singular. These special points are called  \textit{intrinsic}
degenerate critical points of $V(z,\bar{z})$. On the other hand,
the second equation in Eq.~(\ref{degV}) means that some saddle
points of ${\mathcal{W}}(z,\bar{z})$ are mapped into  some
degenerate critical
points of $V(z,\bar{z})$. Note that for a shake of consistency we have assumed that $g_{z\bar{z}}(p_0) >0$.\\
\indent In the model there are two possibilities of Lorentz invariant $N=1$ vacuums
related to the critical point $p_0$ of ${\mathcal{W}}(z,\bar{z})$\footnote{In the following
we mention Lorentz invariant $N=1$ vacuums only as vacuums or ground states.}.
First, the critical point describing Minkowskian spacetime
satisfies the following  conditions
\begin{equation}
      W(z_0) = \frac{dW}{dz}(z_0)=0\:.
\label{Minksing}
\end{equation}
Moreover, it follows that both eigenvalues and determinant in
Eqs.~(\ref{eigenvalW}) and (\ref{eigenvalV}) can be simplified as
follows,
\begin{eqnarray}
\lambda^{\mathcal{W}}_{1,2} &=&  \pm  \, e^{K (p_0)/2M^2_P} \, \left\vert \frac{d^2W}{dz^2}(z_0)\right\vert \nonumber \:,\\
{\textrm{det}}H_{\mathcal{W}} &=& -  \, e^{K (p_0)/M^2_P} \, \left\vert \frac{d^2W}{dz^2}(z_0)\right\vert^2 \:, \\
\lambda^{V}_{1,2} &=& 2
g^{z\bar{z}}(p_0) \, e^{K (p_0)/M^2_P} \, \left\vert \frac{d^2W}{dz^2}(z_0)\right\vert^2 \:,  \nonumber\\
{\textrm{det}}H_V &=& 4\,g^{-2}_{z\bar{z}}(p_0)  \, e^{2K
(p_0)/M^2_P} \, \left\vert \frac{d^2W}{dz^2}(z_0)\right\vert^4
\:.\nonumber
\end{eqnarray}
Non-degeneracy requires
\begin{equation}
 \frac{d^2W}{dz^2}(z_0) \ne 0 \:,\label{Miniso}
 \end{equation}
 which follows that the superpotential $W(z)$ has to be at least quadratic form.
 In other words,  the family of  Minkowskian ground states  satisfying
 Eq.~(\ref{Miniso}) is said to be isolated. Moreover, in these vacuums the possible
 non-degenerate critical points of
${\mathcal{W}}(z,\bar{z})$ are saddles which are mapped into the minima
of the scalar potential, $V(z,\bar{z})$. Both critical points of
${\mathcal{W}}(z,\bar{z})$ and the scalar potential (\ref{V}) will
be degenerate or called non-isolated if Eq.~(\ref{Miniso}) vanishes. An interesting feature
of this case is that these properties do not change if we set $M_P \to +\infty$. This means
that there is an isomorphism between Minkowskian vacuums in local and
global supersymmetric theories for $W(z_0)=0$. 
Finally we want to note that although the first order analysis does not depend on the $U(1)$-connection,
 however in general if we include the higher order terms it does
play a role in Minkowskian vacuums.\\
\indent On the other hand, the critical point describing Anti-de
Sitter (AdS) spacetime has to be obeyed by the following conditions
\begin{equation}
W(z_0)\ne 0 \:, \quad \nabla_z W(p_0)\equiv \frac{dW}{dz}(z_0)+
(K_z(z_0,\bar{z}_0)/M^2_P) W(z_0)=0 \:. \label{AdScon}
\end{equation}
Isolated AdS ground states demand that the function
${\mathcal{W}}(z,\bar{z})$ must not
satisfy Eq.~(\ref{degV}). 
Unlike Minkowskian vacuums, the first order analysis of AdS vacuums does depend on
 the $U(1)$-connection \textit{i.e.}, the term $(K_z(z_0,\bar{z}_0)/M^2_P)$ which is
non-holomorphic, beside superpotential $W(z)$.
To solve such non-holomorphic equation we choose the critical points of the real function
${\mathcal{W}}(z,\bar{z})$ are determined by the critical points of the holomorphic superpotential
$W(z)$, \textit{i.e.}, that
\begin{equation}
        \frac{dW}{dz}(z_0)= 0 \ ,
\end{equation}
which implies that
\begin{equation}
        K_z(z_0,\bar{z}_0) = K_{\bar{z}}(z_0,\bar{z}_0)=0  \:,\label{simply}
\end{equation}
since $W(z_0)\ne 0$. 
This also means that at the ground states the $U(1)$-connection does not appear
in any order analysis. This case will be discussed in section \ref{Models}.

\section{Gradient Flows versus RG Flows}
\label{GRG}

In this section we discuss properties of the gradient flow
equations (\ref{gfe}) and RG flow described by the beta function (\ref{beta})
around critical points (or equilibrium points) of ${\mathcal{W}}(z,\bar{z})$.\\
\indent Let us first consider the gradient flows using dynamical system analysis \cite{dynsis}.
 We say that $p_0$ is an equilibrium point of the gradient flow equation (\ref{gfe}) if
\begin{eqnarray}
z'(p_0) &=& \mp 2 g^{z\bar{z}}
\bar{\partial}_{\bar{z}}{\mathcal{W}}(p_0) = 0 \quad,\nonumber\\
\bar{z}'(p_0) &=& \mp 2 g^{z\bar{z}}
\partial_z{\mathcal{W}}(p_0) = 0 \quad,
\end{eqnarray}
are fulfilled. These follow that the equilibrium points of
Eq.~(\ref{gfe}) are also the critical points of
${\mathcal{W}}(z,\bar{z})$. We then expand Eq.~(\ref{gfe}) near its
equilibrium point up to first order. It turns out that the  first
order expansion matrix has eigenvalues
\begin{equation}
 \Lambda_{1,2}= \mp g^{z\bar{z}}(p_0)\lambda^{\mathcal{W}}_{1,2}
 = \mp \frac{{\mathcal{W}}(p_0)}{M_P^2}- 2 g^{z\bar{z}}(p_0)\lvert \partial^2_z {\mathcal{W}}(p_0)\rvert
 \label{eigenf} \:.
\end{equation}
Stable flow further demands that the two eigenvalues have to be
negative due to Lyapunov theorem. We then find
\begin{equation}
  \lvert \partial^2_z {\mathcal{W}}(p_0)\rvert >
\frac{g_{z\bar{z}}(p_0)}{2M_P^2}{\mathcal{W}}(p_0)\:, \label{node}
\end{equation}
which are stable nodes. Comparing Eq.~(\ref{node}) with Eqs.
(\ref{minV}) and (\ref{saddV}) we see that these flows are flowing
along local minima and the stable directions of saddles of the
scalar potential (\ref{V1}). In other words, the evolution of
domain walls is stable in the context of dynamical system. For
unstable
 flow in the model we have only unstable saddle  since Eq. (\ref{eigenf}) has one possible positive
eigenvalue if the following condition
\begin{equation}
  \lvert \partial^2_z {\mathcal{W}}(p_0)\rvert <
\frac{g_{z\bar{z}}(p_0)}{2M_P^2}{\mathcal{W}}(p_0)\:,
\end{equation}
is satisfied. This condition occurs only at the local maximum of
the scalar potential, namely, the condition (\ref{maxV}). Our
linear analysis fails when the Hessian matrix $H_{\mathcal{W}}$
becomes singular. As discussed in previous section, this occurs at
intrinsic degenerate critical points of the scalar potential
$V(z,\bar{z})$. In other words, the condition (\ref{degW}) is
fulfilled. In dynamical system picture these points could be a
 bifurcation point of the gradient flow equation (\ref{gfe}). Since we have only
a zero eigenvalue, then such bifurcation is called \textit{fold
bifurcation} \cite{dynsis}. These points are assured by at least
one of the higher order terms evaluated at the ground states in
Taylor expansion of Eq. (\ref{gfe}), namely,
\begin{eqnarray}
\partial^n_z \left( g^{z\bar{z}} \partial_z {\mathcal{W}}\right)(p_0) &=&
\sum^n_{q=0} \frac{n!}{(n-q)!}\; \partial^q_z g^{z\bar{z}}(p_0) \;\partial^{n-q+1}_z {\mathcal{W}}(p_0)
\nonumber\\
\partial^m_{\bar{z}}\left( g^{z\bar{z}} \partial_z {\mathcal{W}}\right)(p_0) &=&
\sum^m_{q=0} \frac{m!}{(m-q)!}\; \partial^q_{\bar{z}} g^{z\bar{z}}(p_0) \;
\partial^{m-q}_{\bar{z}}\partial_z {\mathcal{W}}(p_0) \label{highorder}
\end{eqnarray}
for $m \ge 2 , \, n \ge 2$,
\begin{equation}
\partial^m_{\bar{z}}\partial^n_{z} \left( g^{z\bar{z}} \partial_z {\mathcal{W}}\right)(p_0)
= \sum^n_{q=0} \sum^m_{r=0} \frac{n!}{(n-q)!}\; \frac{m!}{(m-r)!}\;
\partial^r_{\bar{z}} \partial^q_z g^{z\bar{z}}(p_0) \; \partial^{m-q}_{\bar{z}}
\partial^{n-q+1}_z {\mathcal{W}}(p_0) \label{highorder1}
\end{equation}
for $m \ge 1 , \, n \ge 1$, and their complex conjugate is have to be non-zero. \\
\indent As discussed in the previous section, in Minkowskian vacuums
the eigenvalue of Eq.~(\ref{eigenf}) can be negative or zero. For
negative value, the flows are guaranteed to be stable flowing
around isolated minimal Minkowskian ground states, while our
linear analysis failed when it vanishes. In AdS cases,
there are two possible flows, namely, stable nodes or unstable
saddles.\\
\indent Now we turn to consider the RG flows. As we have mentioned in section \ref{BPSDW},
this function can also be used to determine the nature of the critical point $p_0$, namely,
it can be interpreted as UV or IR in the CFT picture. To begin, we expand the beta function (\ref{beta})
to first order around $p_0$ and then we have the matrix
\begin{equation}
{\mathcal{U}} \equiv - \left(
\begin{array}{ccc}
  \frac{\partial \beta}{\partial z}(p_0) & & \frac{\partial \bar{\beta}}{\partial z}(p_0) \\
  & & \\
  \frac{\partial \beta}{\partial \bar{z}}(p_0) & & \frac{\partial \bar{\beta}}{\partial \bar{z}  }(p_0) \\
\end{array}
\right)\:,
\end{equation}
whose eigenvalues are
\begin{equation}
\lambda^{\mathcal{U}}_{1,2} = \frac{2g^{z\bar{z}}(p_0)}{{\mathcal{W}}(p_0)}\, \lambda^{\mathcal{W}}_{1,2}
= \frac{2}{M_P^2} \pm  \frac{4g^{z\bar{z}}(p_0) }{{\mathcal{W}}(p_0)}\,\lvert \partial^2_z {\mathcal{W}}(p_0)\rvert\:\,.
\label{eigenU}
\end{equation}
Let us choose a model where the UV region is $u \to + \infty$ as $a \to + \infty$,
while the IR region is $u \to - \infty$ as $a \to 0$. First we consider the UV critical points.
In UV region at least one of the eigenvalues (\ref{eigenU}) should be positive. Since
it is possible to have zero and a negative eigenvalue as the parameters in
${\mathcal{W}}(z,\bar{z})$ vary, namely, in the direction of
\begin{equation}
\lambda^{\mathcal{U}}_2 = \frac{2}{M_P^2} - \frac{4g^{z\bar{z}}(p_0) }{{\mathcal{W}}(p_0)}\,
\lvert \partial^2_z {\mathcal{W}}(p_0)\rvert\:\,,
\label{eigenUneg}
\end{equation}
the RG flows fail to depart the UV region in this direction. Thus, this flow is stable along
\begin{equation}
\lambda^{\mathcal{U}}_1 = \frac{2}{M_P^2} +
\frac{4g^{z\bar{z}}(p_0) }{{\mathcal{W}}(p_0)}\, \lvert
\partial^2_z {\mathcal{W}}(p_0)\rvert\:\,. \label{eigenUpos}
\end{equation}
Correspondingly, the local maxima of the scalar potential $V(z,\bar{z})$ are changing into saddles or
 local minima in this UV region. Then the stability of gradient flows is changing, namely,
 from unstable saddles into stable nodes.  In addition, it is easy to see that bifurcation point
 of gradient flows does exist in UV region. On the other hand, in IR region the RG flow approaches
 a critical point  in the direction where the eigenvalue of Eq.~(\ref{eigenUneg}) is negative.
 However, it is failed when  Eq. (\ref{eigenUneg}) vanishes and then becomes positive.
 Thus, IR critical points vanish as the parameters varying.
 In other words, there is no IR local maximum of the scalar potential and the gradient flows do not exist
 around this point in the model. Moreover, we have no bifurcation
 point of gradient flows in this region.

\section{Models}
\label{Models} \indent We organize this section into three parts.
First we consider a model where the superpotential are algebraic
polynomials, namely, linear, quadratic, and cubic polynomials.
Then in the second part  a model with harmonic superpotential will
be discussed. The last part will be a model with elliptic
superpotential which describes Riemann surfaces of genus one. For
the rest of the paper we take the K\"ahler potential to be
\begin{equation}
 K(z,\bar{z}) = |z-z_0|^2  \:, \label{Kexa}
\end{equation}
where the parameter $z_0$ is chosen such that in vacuums it satisfies Eq.~(\ref{simply}).
Then this choice further simplifies Eqs.~(\ref{eigenvalW}) and (\ref{eigenvalV}) into
\begin{eqnarray}
\lambda^{\mathcal{W}}_{1,2} &=& \frac{|W(z_0)|}{M_P^2}
 \pm  \left\vert \frac{d^2W}{dz^2}(z_0)\right\vert \nonumber \:,\\
 {\textrm{det}}H_{\mathcal{W}} &=&  \frac{|W(z_0)|^2}{M_P^4}
 -  \left\vert \frac{d^2W}{dz^2}(z_0)\right\vert^2 \:,\label{SimpVW} \\
 \lambda^{V}_{1,2} &=& 2 \left( \left\vert \frac{d^2W}{dz^2}(z_0)\right\vert^2
  - \frac{2|W(z_0)|^2}{M_P^4}  \pm \,  \frac{|W(z_0)|}{M_P^2}  \:
  \left\vert \frac{d^2W}{dz^2}(z_0)\right\vert \right)  \:,\nonumber\\
&& \nonumber\\
 {\textrm{det}}H_V &=& 4 \left( \frac{|W(z_0)|^2}{M_P^4}
 -  \left\vert \frac{d^2W}{dz^2}(z_0)\right\vert^2 \right)\, \: \left( \frac{4|W(z_0)|^2}{M_P^4}
 - \left\vert \frac{d^2W}{dz^2}(z_0)\right\vert^2 \right)\nonumber\:.
\end{eqnarray}
We see that in this model the dynamics of the walls are completely determined by
the superpotential $W(z)$. As mentioned in section \ref{CPWV} Minkowskian vacuums are
stable and isolated for at least quadratic form in both local and global $N=1$
supersymmetric theories.

\subsection{A Model with $W(z)$ as Algebraic Polynomials}
\label{PSW}
In this section we consider a model in which the holomorphic superpotential has the
following form
\begin{equation}
  W(z) = \sum^{\mathcal{N}}_{n=0} a_n z^n \:,
\end{equation}
with $a_n \:, n=0,..,{\mathcal{N}}$ are complex. In particular we will focus, as examples, on
${\mathcal{N}}=0, 1, 2$.\\
\indent Let us first consider two simplest models, namely,  ${\mathcal{N}}=0$ and linear model
${\mathcal{N}}=1$. In these models we have
\begin{eqnarray}
\lambda^{\mathcal{W}}_{1,2} &=& \frac{|a_0|}{M_P^2}\:, \quad \Lambda_{1,2} = \mp \frac{|a_0|}{M_P^2} \:,\nonumber\\
\lambda^{V}_{1,2} &=& -4 \frac{|a_0|^2}{M_P^4}\:, \quad V(z_0,\bar{z}_0) = - \frac{3|a_0|^2}{M_P^2} \:,
\end{eqnarray}
since the critical points are trivial for ${\mathcal{N}}=0$ and
$a_1 =0$ for ${\mathcal{N}}=1$. AdS vacuums occur if $a_0 \ne 0$
which are unstable and isolated, while  these models admit trivial
Minkowskian vacuums if $a_0 = 0$ which is not a bifurcation point.
Furthermore, in both models we have only UV vacuums. Note that these
two models are degenerate in the sense that
they have the same properties at the ground states.\\
\indent Now we turn to consider the case for ${\mathcal{N}}=2$. In the model we have
\begin{eqnarray}
\lambda^{\mathcal{W}}_{1,2} &=& \frac{|D|}{4|a_2|M_P^2} \pm 4 |a_2| \:,\nonumber\\
\Lambda_{1,2} &=&  \mp \frac{|D|}{4|a_2|M_P^2} - 4 |a_2| \:,\nonumber\\
\lambda^{V}_{1,2} &=& 8 |a_2|^2 - \frac{|D|^2}{16|a_2|^2M_P^4} \pm \frac{|D|}{2M_P^2} \:,\\
V(z_0,\bar{z}_0) &=& - \frac{3|D|^2}{16|a_2|^2M_P^2} \:, \nonumber
\end{eqnarray}
where $D= a_1^2 - 4 a_2\, a_0$. Clearly, Minkowskian vacuums emerge if $D = 0$ and $a_2 \ne 0$
which are stable and isolated in UV or IR regions. The case where $D = 0$ and $a_2 = 0$ one should regain
the previous models which are AdS for $a_0 \ne 0$. Particularly, the analysis is blown up for $D \ne 0$ and $a_2 = 0$
in UV region.

\subsection{A Model with Harmonic $W(z)$}
This section is devoted to discuss properties of a model where the superpotential $W(z)$
has the harmonic form\footnote{This harmonic form of superpotential with gravitational correction
has been discussed in Ref.~\cite{BPSEto}.}
\begin{equation}
  W(z) = A_0 \,{\mathrm{sin}}(kz) \:,
\end{equation}
where $A_0$ and $k$ are real. Furthermore, by defining $z \equiv x + {\mathrm{i}}y$, the critical points are
\begin{equation}
  y = 0 \:,\: x= (n+ \frac{1}{2})\frac{\pi}{k} \:, \quad n = 0, \pm 1, \pm 2, ....
\end{equation}
Then the eigenvalues in Eqs.~(\ref{SimpVW}) and (\ref{eigenvalW})
and the scalar potential (\ref{V1}) becomes
\begin{eqnarray}
\lambda^{\mathcal{W}}_{1,2} &=& \left( \frac{1}{M_P^2} \pm k^2 \right)|A_0| \:,\nonumber\\
\Lambda_{1,2} &=&  \left( \mp \frac{1}{M_P^2} - k^2 \right)|A_0| \:,\nonumber\\
\lambda^{V}_{1,2} &=& 2|A_0|^2 \left( k^4- \frac{2}{M_P^4} \pm \frac{k^2}{M_P^2}\right) \:,\\
V(z_0,\bar{z}_0) &=& - \frac{3|A_0|^2}{M_P^2} \:. \nonumber
\end{eqnarray}
In this model we have only nontrivial AdS vacuums and the possible bifurcation point happens at $k = \pm \frac{1}{M_P}$.

\subsection{A Model with Elliptic $W(z)$}
\label{EW}
In this section we give example of a model where the holomorphic
superpotential $W(z)$ can be viewed as Weierstrass function, \cite{FK}
\footnote{This function has been studied in the context of dilaton domain walls (Ref.~\cite{GravStab}).}
\begin{equation}
\left(\frac{dW(z;\tau)}{dz}\right)^2
=\big(W(z;\tau)-e_1(\tau)\big)\big(W(z;\tau)-e_2(\tau)\big)\big(W(z;\tau)-e_3(\tau)\big)
\:, \label{weier}
\end{equation}
where the roots $e_1, e_2, e_3$ are complex and moduli
$\tau$-dependent. These can be expressed in terms of theta
functions $\vartheta(0;\tau)$,
\begin{eqnarray}
 e_1(\tau) &=&  \frac{\pi^2}{3} \big(\vartheta^4(0;\tau)+ \vartheta^4_{01}(0;\tau)\big)\:,\nonumber\\
e_2(\tau) &=&  -\frac{\pi^2}{3} \big(\vartheta^4(0;\tau)+ \vartheta^4_{10}(0;\tau) \big)\:,\\
e_3(\tau) &=&  - \big(e_1(\tau)+ e_2(\tau) \big)\:,\nonumber
\end{eqnarray}
where
\begin{eqnarray}
\vartheta(z;\tau)  &=&   \sum^{+\infty}_{n=-\infty}q^{n^2 /2}\: e^{2\pi{\mathrm{i}}n z}\:,\nonumber\\
 \vartheta_{01}(z;\tau) &=&  \sum^{+\infty}_{n=-\infty} (-1)^n \; q^{n^2/2} \: e^{2\pi{\mathrm{i}}n z}\:,\\
 \vartheta_{10}(z;\tau)&=&   \;  \sum^{+\infty}_{n=-\infty}
  \; q^{\frac{1}{2}(n + \frac{1}{2} )^2 }\: e^{(2n+1)\pi{\mathrm{i}}z} \:,\nonumber
\end{eqnarray}
with $q \equiv e^{2\pi {\mathrm{i}} \tau}$. Furthermore the superpotential $W(z;\tau)$ satisfying Eq.~(\ref{weier})
can also be written down in  terms of theta
functions $\vartheta(z;\tau)$,
\begin{equation}
W(z;\tau) = \pi^2 \, \vartheta^2(0;\tau) \, \vartheta^2_{10}(0;\tau) \, \frac{\vartheta^2_{01}(z;\tau)}
{\vartheta^2_{11}(z;\tau)} + e_2(\tau) \ ,
\label{weier1}
\end{equation}
where
\begin{equation}
\vartheta_{11}(z;\tau) =   \;  \sum^{+\infty}_{n=-\infty}
  \; (-1)^n \,q^{\frac{1}{2}(n + \frac{1}{2} )^2 }\: e^{(2n+1)\pi{\mathrm{i}}z} \:.
\end{equation}
\indent Now we recall about critical points of ${\mathcal{W}}(z,\bar{z})$.
For the case at hand we find that if $p_0$ is a critical point  of
${\mathcal{W}}(z,\bar{z})$, then we have
\begin{equation}
     W(z_0;\tau)= e_l(\tau) \:, \qquad l =1,2,3 .
\end{equation}
Then the scalar potential (\ref{V}) becomes
\begin{equation}
 V(\tau,\bar{\tau}) = -  \frac{3\vert e_l(\tau) \vert^2 }{M^2_P} \, \,
 \qquad,\;{(\mathrm{no \: summation})},
\end{equation}
which is the cosmological constant of the model. However our
analysis (\ref{SimpVW}) becomes ill defined since it diverges at
$p_0$. These singularities cannot be removed and can be found in
another part of the upper-half plane by employing modular
transformation with respect to $\tau$. This means that our
analysis fails for this function near its
critical points.\\

\section{Conclusions}
\label{Conclu}
In this paper we have studied general properties of $N=1$ vacuums of BPS equations and
the scalar potential in four dimensional $N=1$ supergravity coupled to a chiral multiplet.
Firstly, our analysis on the real function ${\mathcal{W}}(z,\bar{z})$ shows that this function
admits three types of critical points, namely, local minima, saddles, and degenerate critical points.
The local minima are mapped into local maxima of the scalar potential, while the saddles
are mapped into local minima or saddles of the scalar potential. Moreover, degeneracy of
${\mathcal{W}}(z,\bar{z})$ also implies degeneracy of the scalar potential. These points are
called intrinsic degenerate critical points of the scalar potential.\\
\indent Secondly, using dynamical system analysis on gradient flow equations we obtain
 that in this model these flows are unstable near local maximum of
the scalar potential, whereas they are stable near saddles, local
minimum, and degenerate critical points of the scalar potential.
Furthermore, we have showed that in this model it is possible to
have a bifurcation point occurring near intrinsic degenerate
critical points of the scalar potential. Thirdly, to check the
existence of such flows near critical points one has to perform
the analysis on the beta function. Our conclusion is that IR local
maxima does not exist  in the model
which further tells us that no gradient flows exist around these points in the region. \\
\indent Finally, we have considered three models where the ground
states are completely controlled by the superpotential. In the
model with algebraic polynomials we find that there is degeneracy
between the two models, namely, constant and linear superpotentials.
These simple models admit nontrivial unstable and isolated AdS
vacuums in UV region. For the case of quadratic polynomial the
situation is more complicated. In particular, our analysis diverges
for $D \ne 0$ and $a_2 = 0$ in UV region.\\
\indent In the model with harmonic superpotential we have periodic
critical points and then the nontrivial ground states are AdS.
Lastly,  the model with elliptic superpotential has also been
considered . However, our analysis becomes divergence near
critical points in this model.

\appendix
\section{Convention and Notation}
The purpose of this appendix is to collect  our conventions in
this paper. The spacetime metric is taken to have the signature
$(+,-,-,-)$ while the Ricci scalar is defined to be
$R=g^{\mu\nu}\Big[\frac{1}{2}\partial_{\rho}\Gamma^{\rho}_{\mu\nu}
-\partial_{\nu}\Gamma^{\rho}_{\mu\rho}+\Gamma^{\sigma}_{\mu\nu}\Gamma^{\rho}_{\sigma\rho}\Big]
+\frac{1}{2}\partial_{\rho}\Big[g^{\mu\nu}\Gamma^{\rho}_{\mu\nu}\Big]$.
The Christoffel symbol is given by
$\Gamma^{\mu}_{\nu\rho}=\frac{1}{2}g^{\mu\sigma}(\partial_{\nu}g_{\rho\sigma}+\partial_{\rho}g_{\nu\sigma}-\partial_{\sigma}g_{\nu\rho})$
where $g_{\mu\nu}$ is the spacetime metric.\\

The following indices are given:\\

\begin{tabular}{r @{\hspace{2.5 cm}}  l }
$\overline{\mu},\overline{\nu} = 0,1,2$, & label curved three dimensional spacetime indices \\
\\
$\overline{a}, \overline{b} = 0,1,2$, & label flat three dimensional spacetime indices \\
\\
$\mu, \nu = 0,...,3$, & label curved four dimensional spacetime indices \\
\\
$a, b = 0,...,3$, & label flat four dimensional spacetime indices \\
\\
\\
\\
\\
\\
\\
\end{tabular}

\section{Hessian Matrix}
\label{Hess}
\indent Let us first consider an arbitrary (real)
$C^\infty$-function $f(z,\bar{z})$. A point $p_0 =
(z_0,\bar{z}_0)$ is said to be a critical point of $f(z,\bar{z})$
if the following conditions
\begin{equation}
\frac{\partial f}{\partial z}(p_0)=0 \:,\quad \frac{\partial
f}{\partial \bar{z}}(p_0)=0 \:,
\end{equation}
hold. Furthermore, the point $p_0$ is a non-degenerate critical
point of $f(z,\bar{z})$ if the Hessian matrix
\begin{equation}
H_f  \equiv 2 \left(
\begin{array}{ccc}
  \frac{\partial^2 f}{\partial z \partial \bar{z}}(p_0) & & \frac{\partial^2 f}{\partial z^2}(p_0) \\
  & & \\
  \frac{\partial^2 f}{\partial \bar{z}^2}(p_0) & & \frac{\partial^2 f}{\partial \bar{z} \partial z }(p_0) \\
\end{array}
\right)
\end{equation}
is non-singular, \textit{i.e.},
\begin{equation}
 {\textrm{det}}\,H_f = 4 \left[ \left(
\frac{\partial^2 f}{\partial z \partial \bar{z}}(p_0)\right)^2
-\frac{\partial^2 f}{\partial z^2}(p_0)\frac{\partial^2
f}{\partial \bar{z}^2}(p_0) \right] \ne 0 \: .
\end{equation}
The eigenvalues of the Hessian matrix $H_f$ are given by
\begin{eqnarray}
\lambda^f_1 &=& \frac{1}{2} \left( {\textrm{tr}}H_f +
\sqrt{\left({\textrm{tr}}H_f \right)^2-4\, {\textrm{det}}H_f
}\right) \nonumber \:,\\
\lambda^f_2 &=& \frac{1}{2} \left( {\textrm{tr}}H_f -
\sqrt{\left({\textrm{tr}}H_f \right)^2-4\, {\textrm{det}}H_f
}\right) \:.\label{eigenval}
\end{eqnarray}
\indent Now we can characterize the critical point $p_0$ of the
function $f$ using the eigenvalues defined in Eq.~(\ref{eigenval}) as
follows:
\begin{itemize}
\item[1.] If two eigenvalues are positive, \textit{i.e.},
$\lambda^f_1 > 0$ and $\lambda^f_2 > 0$, then $p_0$ is a local
minimum. This implies
\begin{eqnarray}
 {\textrm{tr}}H_f &>& 0 \nonumber \:,\\
 {\textrm{det}}H_f &>& 0 \:.
\end{eqnarray}
\item[2.] If two eigenvalues are negative, \textit{i.e.},
$\lambda^f_1 < 0$ and $\lambda^f_2 < 0$, then $p_0$ is a local
maximum which satisfies
\begin{eqnarray}
 {\textrm{tr}}H_f &<& 0 \nonumber \:,\\
 {\textrm{det}}H_f &>& 0 \:.
\end{eqnarray}
\item[3.] If we have $\lambda^f_1 > 0$ and $\lambda^f_2 < 0$ or
vice versa, then $p_0$ is a saddle point which satisfies
\begin{equation}
  {\textrm{det}}H_f < 0 \:.
\end{equation}
\item[4.] Finally if at least one eigenvalue vanishes, then $p_0$
is a degenerate critical point. This means that we have
\begin{equation}
  {\textrm{det}} H_f = 0 \:.
\end{equation}
\end{itemize}

\newpage

\hspace{-0.6 cm}{\Large \bf Acknowledgement}
\\
\vskip 0.15truecm \hspace{-0.6 cm}We acknowledge  H. Alatas and
J.M. Tuwankotta for nice discussion about bifurcation theory. We
would like to thank M. Cvetic for notifying Refs.~\cite{BPSDW} and
\cite{RevBPSDW} and K. Yamamoto for informing us
Ref.~\cite{CDGKL}. This work is supported by Riset ITB 2005 No.
0004/K01.03.2/PL2.1.5/I/2006.


\vskip 0.15truecm
\begin{thebibliography}{99}
%
\bibitem{AdSCFT}
For a review see, for example: \\
O. Aharony, S. S. Gubser, J. Maldacena, H. Ooguri, and Y. Oz,
Phys. Rep. {\bf 323} 183 (2000), and references therein, e-print:hep-th/9905111.
\bibitem{AC}
A. Celi,
Ph.D. Thesis Universita d'Milano,  and references therein, e-print:hep-th/0405283.
\bibitem{BPSDW}
M. Cvetic, S. Griffies and S.J. Rey,
Nucl. Phys. B {\bf 381} 301 (1992), e-print:hep-th/9201007.
\bibitem{RevBPSDW}
M. Cvetic and H.H. Soleng,
  Phys. Rep.  {\bf 282} 159 (1997), e-print:hep-th/9604090.
\bibitem{DDW}
J. Sonner and P.K. Townsend,
Class. Quantum Grav. {\bf 23}, 441 (2006), e-print:hep-th/0510115.
\bibitem{CDGKL}
A. Ceresole, G. Dall'Agata, A. Girvayets, R. Kallosh, and A.
Linde, \textit{Domain walls, near-BPS bubbles and probabilities in
the landscape}, Phys. Rev. {\bf D74} (2006) 086010,
hep-th/0605266.
\bibitem{susy}
J. Wess and J. Bagger,
\textit{Supersymmetry and Supergravity}, 2nd ed. (
Princeton University Press, Princeton 1992).
R. D'Auria and S. Ferrara ,
JHEP {\bf 0105}, 034 (2001), e-print:hep-th/0103153.
L. Andrianopoli, R. D'Auria, and S. Ferrara,
JHEP {\bf 0203},  025 (2002), e-print:hep-th/0110277.
\bibitem{DWc}
D.Z. Freedman, S.S. Gubser, K. Pilch and N.P. Warner,
\textit{Renormalization group flows from holography-supersymmetry
and a c-theorem}, Adv. Theor. Math. Phys. {\bf 3} (1999)
363, hep-th/9904017.
\bibitem{CDKV}
A. Ceresole, G. Dall'Agata, R. Kallosh, and A. van Proeyen, \textit{Hypermultiplets, Domain Walls and
Supersymmetric Attractor},  Phys. Rev. {\bf D64} (2001) 104006, hep-th/0104056.
\bibitem{calculus}
W. Kaplan, \textit{Advanced Calculus}, (2 ed.), Addison Wesley
(1973).
\bibitem{dynsis}
Y. A. Kuznetsov, \textit{Elements of Applied Bifurcation Theory},
Springer-Verlag New York (1998).
\bibitem{BPSEto}
M. Eto, N. Maru, N. Sakai, and T. Sakata, \textit{Exactly solved BPS wall and winding number
in $N=1$ Supergravity},  Phys. Lett. {\bf B553} (2003) 87, hep-th/0208127.
\bibitem{FK}
H. M. Farkas and I. Kra, \textit{Riemann Surfaces},
Springer-Verlag New York (1980).\\
N. Koblitz, \textit{Introduction to Elliptic Curves and Modular
Forms}, Springer-Verlag New York (1993).
\bibitem{GravStab}
K. Skenderis and P.K. Townsend, \textit{Gravitational Stability and Renormalization Group Flow},
Phys. Lett. {\bf B468} (1999) 46, hep-th/9909070.


\end{thebibliography}
\end{document}